# Superconductivity in Weyl Semimetal Candidate $MoTe_2$

Yanpeng Qi[1], Pavel G. Naumov[1], Mazhar N. Ali[2], Catherine R. Rajamathi[1], Walter Schnelle[1], Oleg Barkalov[1], Michael Hanfland[3], Shu-Chun Wu[1], Chandra Shekhar[1], Yan Sun[1], Vicky Süß[1], Marcus Schmidt[1], Ulrich Schwarz[1], Eckhard Pippel[4], Peter Werner[4], Reinald Hillebrand[4], Tobias Förster[5], Erik Kampert[5], Stuart Parkin[4], R. J. Cava[2], Claudia Felser[1], Binghai Yan[1,6]*, Sergey A. Medvedev[1]*

[1]Max Planck Institute for Chemical Physics of Solids, 01187 Dresden, Germany

[2]Department of Chemistry, Princeton University, Princeton, New Jersey 08544, USA

[3]European Synchrotron Radiation Facility, BP 220, Grenoble 38043, France

[4]Max Planck Institute of Microstructure Physics, 06120 Halle, Germany

[5]Dresden High Magnetic Field Laboratory (HLD-EMFL), Helmholtz-Zentrum Dresden-Rossendorf, 01328 Dresden, Germany

[6]Max Planck Institute for the Physics of Complex Systems, 01187 Dresden, Germany

* E-mail: Yan@cpfs.mpg.de, Sergiy.Medvediev@cpfs.mpg.de

**Abstract**

Transition metal dichalcogenides have attracted research interest over the last few decades due to their interesting structural chemistry, unusual electronic properties, rich intercalation chemistry and wide spectrum of potential applications. Despite the fact that the majority of related research focuses on semiconducting transition-metal dichalcogenides e.g., $MoS_2$, recently discovered unexpected properties of $WTe_2$ are provoking strong interest in semimetallic transition metal dichalcogenides featuring large magnetoresistance, pressure-driven superconductivity, and Weyl semimetal state. We investigate the sister compound of $WTe_2$, $MoTe_2$, predicted to be a Weyl semimetal and a quantum spin Hall insulator in bulk and monolayer form, respectively. We find that bulk $MoTe_2$ exhibits superconductivity with a transition temperature of 0.10 K and application of external pressure dramatically enhances the transition temperature up to maximum value of 8.2 K at 11.7 GPa. Observed dome-shaped superconductivity phase diagram provides insights into the interplay between superconductivity and topological physics.

Transition-metal dichalcogenides (TMDs) have attracted tremendous attention due to their rich physics and promising potential applications[1-11]. TMDs share the same formula, $MX_2$, where M is a transition metal (e.g., Mo or W) and X is a chalcogenide atom (S, Se, and Te). These compounds typically crystallize in many structures, including 2H-, 1T-, 1T'-, and $T_d$-type lattices. The most common structure is the 2H phase, where M atoms are trigonal-prismatically coordinated by the chalcogenide atoms. These planes then stack upon one other with van der Waals gaps inbetween. In contrast, the 1T structure corresponds to octahedral coordination of M. The 1T' phase is a monoclinic lattice that can be interpreted as a distortion of the 1T phase by the formation of in-plane M-M bonds, resulting in a pseudo-hexagonal layer with zigzag metal chains. Finally, the $T_d$ phase is very similar to the 1T' phase, but the layers stack in a direct fashion, resulting in a higher-symmetry orthorhombic structure. Depending on the synthesis technique, the same composition of $MX_2$ can crystallize in a variety of structures with very different electronic properties. For example, $MoTe_2$ exists in 2H, 1T', and $T_d$ structures[12-14], while $WTe_2$ has commonly been observed in the $T_d$ structure[15]. The 2H and 1T compounds are primarily semiconducting, whereas the 1T' and $T_d$ compounds are typically semimetallic.

Very recently, semimetallic TMDs have attracted considerable attention because of the discovery of salient quantum phenomena. For instance, $T_d$-$WTe_2$ has been found to exhibit an extremely large magnetoresistance[16,17], pressure (*P*)-driven superconductivity (highest resistive transition temperature $T_c \approx 7$ K at 16.8 GPa)[18,19], and a large and linear Nernst effect[20]. Further, this material has been theorized to constitute the first example of a type-II Weyl semimetal[21]. Moreover, the 1T'-$MX_2$ monolayer has been predicted to be a 2D topological insulator [6].

The discovery of superconductivity in $WTe_2$ is apparently contradictory to previous theoretical predictions[22], which claim that 2H TMDs may become superconducting at high *P*, but the 1T' phases will not. Thus the investigation of other TMDs for the appearance of superconductivity under pressure is of big interest. Molybdenum ditelluride ($MoTe_2$) is unique among the TMDs since it is the only material that can be grown in both 2H and 1T' forms, allowing for direct examination of this theory. If superconductivity exists in 1T'-$MoTe_2$, it may allow the topological edge states to also become superconducting, because of the proximity effect in a bulk superconductor. This would open up a new platform for the study of topological superconductivity, which has potential application in quantum computation[23]. Regarding the recently anticipated Weyl semimetal (WSM) phase in $MoTe_2$ (Ref. 24), discovery of superconductivity may introduce a new pathway for the exploration of topological superconductivity[25-27] along with emergent space-time supersymmetry [28].

Here, we report on the transport properties of the 2H, 1T', and $T_d$ polytypes of $MoTe_2$ under various applied *P*. We find that $T_d$-$MoTe_2$ exhibits superconductivity with $T_c$ = 0.10 K, according to electrical resistivity ($\rho$) measurements. Application of relatively low pressures below 1 GPa dramatically enhances the $T_c$, and a dome-shaped $T_c$-$P$ phase diagram is observed with maximum $T_c$ = 8.2 K at 11.7 GPa; this is approximately 80 times larger than the ambient pressure value. In contrast, we do not observe any traces of superconductivity in the 2H phase, even when it becomes metallic under *P*. We assume that the extreme sensitivity of the superconductivity to *P* is a consequence of the unique electronic structure. Thus, $MoTe_2$ presents the opportunity to study the interaction of topological physics and superconductivity in a bulk material.

## Results

### Structure and transport properties at ambient pressure

Prior physical properties measurements, synthesized 1T'-$MoTe_2$ samples were characterized using single-crystal x-ray diffraction (SXRD) and high-angle annular dark-field scanning transmission electron microscopy (HAADF-STEM). The atomic arrangement of the 1T' structure was determined using high-resolution HAADF-STEM images and diffraction patterns, as shown in Fig. 1a, b and Supplementary Fig. 1a, b. The crystal structures of 1T' and $T_d$-$MoTe_2$ are sketched in Fig. 1c. At room temperature, the crystals exhibit the expected monoclinic 1T'-$MoTe_2$ structure, while the SXRD measurements at 120 K indicate a transition into the orthorhombic $T_d$ structure. 1T'-$MoTe_2$ structure crystallizes in the $P2_1/m$ space group with lattice parameters of $a$ = 6.320 Å, $b$ = 3.469 Å, $c$ = 13.86 Å, and $\beta$ = 93.917º; these results are consistent with the previously reported structure[12]. The Raman spectra at ambient $P$ contain two characteristic peaks (Supplementary Fig. 1c), which are due to the $A_g$ and $B_g$ vibrational modes of the 1T'-$MoTe_2$ structure; this is also in agreement with a previous report[29]. A full structural solution was obtained for the orthorhombic $T_d$ phase at 120 K, the refined parameters are given in Supplementary Tables 1 and 2.

Temperature dependence of electrical resistivity of $MoTe_2$ down to a minimum temperature of $T_{min}$ = 0.08 K at ambient pressure is presented in Fig. 2. In contrast to the 2H phase, which displays semiconducting behavior, 1T'-$MoTe_2$ is semimetallic in nature. At zero field, the room-temperature resistivity is $\rho = 1.0 \times 10^{-5}\ \Omega$ m, which decreases to $2.8 \times 10^{-7}\ \Omega$ m at 0.5 K, yielding a residual resistance ratio (RRR) ≈ 36. At $T \approx 250$ K an anomaly with thermal hysteresis (Fig. 2a, inset) is observed, which is associated with the first-order structural phase transition from the 1T' to the $T_d$ polytype[14,30]. A range of magneto-transport properties has been measured at zero pressure on our $MoTe_2$ crystals (Supplementary Figs. 2-4 and Supplementary Note 1). From Hall effect measurements, $MoTe_2$ shows dominant electron-type transport. Within a single-band model the electron concentration $n_e$ is estimated to 5 x $10^{19}$ $cm^{-3}$ at 2 K and 8 x $10^{20}$ $cm^{-3}$ at 300 K (Supplementary Fig. 2), which is close to reported value[29]. In addition, $T_d$-$MoTe_2$ gradually becomes superconducting below $T \sim 0.3$ K (the onset of transition), while zero resistance is observed at $T_c$ = 0.10 K (Fig. 2b). Note that, although potential superconductivity at ~ 0.3 K in $MoTe_2$ has been briefly mentioned in the literature[31], no related data have been published.

**1T' – $T_d$ structural transition under pressure**

It is well known that high pressure can effectively modify lattice structures and the corresponding electronic states in a systematic fashion. Hence, we measured $\rho(T)$ for the same 1T'-$MoTe_2$ single crystal at various pressure values $P$ (Fig. 3). Fig. 3a shows the typical $\rho(T)$ curves for $P$ up to 34.9 GPa. For increasing $P$, the metallic characteristic increases and $\rho$ decreases over the entire temperature range. At low pressures, resistance curves exhibit an anomaly at a temperature $T_s$, associated with the monoclinic 1T' – orthorhombic $T_d$ structural phase transition similarly to the ambient pressure data. With pressure increase, the resistivity anomaly becomes less pronounced whereas the temperature of anomaly $T_s$ is significantly shifted to lower $T$ and disappears completely above 4 GPa. Thus, the application of $P$ tends to stabilize the monoclinic phase. In addition, the Raman spectra recorded at room temperature under different pressures (Fig. 4a) contain only two characteristic peaks for the 1T′-structure $A_g$ and $B_g$ modes[29]. The frequencies of both vibrational modes increase gradually with no discontinuities as $P$ increases (Fig. 4b) indicating the absence of major structural phase transition in the whole studied pressure range at room temperature. SXRD data (Fig. 4c and Supplementary Fig. 5) also indicate that application of pressure stabilizes the monoclinic 1T' structure. Increase of $P$ at room temperature results in enhancement of monoclinic distortion (increase of the monoclinic angle $\beta$). In an isothermal run at 135 K the reversible orthorhombic $T_d$ to monoclinic 1T' transition is observed at ≈ 0.8 GPa (≈ 0.4 GPa) at pressure increase (decrease) (Fig. 4c). Thus, application of $P$ well below 1 GPa decreases the temperature of structural transition to below 135 K. Furthermore, at $P$ ≈1.5 GPa, the 1T' structure remains stable down to at least 80 K. The quantitative discrepancy in the $T_s$ values derived from structural and resistivity data is most likely due to nonhydrostatic pressure conditions in the resistivity measurements, and the thermal hysteresis since the resistivity curves are recorded with increasing temperatures.

The stability of $MoTe_2$ in different phases can be explained using total energy calculations within density-functional theory (DFT). The optimized lattice constants are very close to experimental values for both two phases, as shown in Supplementary Fig. 6 and Supplementary Table 3. After evaluating the total energies of the two phases at ambient pressure, we found that the $T_d$ phase exhibits slightly lower energy (0.5 meV per formula unit) than the 1T' phase. This is consistent with the fact that the low- and high-*T* phases are $T_d$ and 1T', respectively, without external pressure. As the 1T' phase can be obtained by sliding between layers of the $T_d$ phase, the former exhibits a slightly smaller equilibrium volume than the latter, as also revealed from the lattice parameters measured via SXRD. As illustrated by the energy-volume profile in Fig. 1d, external pressure will stabilize the 1T' phase with the smaller volume (and correspondingly higher density) by increasing the shift between neighboring layers.

**The dome-shaped superconductivity behavior**

Our pressure studies have revealed that the $T_c$ is very sensitive to pressure. That is, $T_c$ increases dramatically to 5 K at relatively low pressures below 1 GPa, before beginning a slower increase to a maximum $T_c$ of 8.2 K at 11.7 GPa (Figs. 3b, 5). Beyond this pressure, $T_c$ decreases and no superconductivity with $T_c > 1.5$ K is found at $P > 34.9$ GPa (Fig. 3c). Remarkably, the drastic increase of $T_c$ at low pressures is associated with sharp decrease of temperature of 1T' - $T_d$ structural phase transition $T_s$. Subsequently at higher pressures, $T_c$ still increases to its maximum value with increasing $P$ but with significantly lower rate. Our findings demonstrate that the strong enhancement of $T_c$ under high $P$ is associated with suppression of the 1T' - $T_d$ structural phase transition. All the characteristic temperatures in the above experimental results are summarized in the *T-P* phase diagram in Fig. 5. A dome-shaped superconducting phase boundary is obtained for $MoTe_2$, with a sharp slope towards the zero-$P$ end of the diagram.

The bulk character of the superconductivity is confirmed by observations of the magnetic shielding effect in the low pressure range and at 7.5 GPa (Supplementary Fig. 7). The onset temperatures of the diamagnetism are consistent with that of the resistivity drop and confirm the drastic increase of $T_c$ in the low pressure range (Fig. 5). Further, we conducted resistivity measurements in the vicinity of $T_c$ for various external magnetic fields. As can be seen in Fig. 3d, the zero-resistance-point $T_c$ under $P$ = 11.2 GPa is gradually suppressed with increasing field. Deviating from the Werthamer-Helfand-Hohenberg theory based on the single-band model, the upper critical field, $H_{c2}(T)$, of $MoTe_2$ has a positive curvature close to $T_c$ ($H$ = 0), as shown in Fig. 3e. This is similar to the behaviors of both $WTe_2$ (Ref. 18) and $NbSe_2$ (Ref. 32). The experimental $H_{c2}(T)$ data can be described within the entire $T/T_c$ range by the expression $H_{c2}(T) = H_{c2}^{*} (1 - T/T_c)^{1+\alpha}$ (Refs. 18, 33). The fitting parameter $H_{c2}^{*} = 4.0$ T can be considered as the upper limit for the upper critical field $H_{c2}(0)$, which yields a Ginzburg–Landau coherence length $\xi_{GL}(0)$ of ~9 nm. The corresponding data obtained at $P$ = 1.1 GPa is also shown in Fig. 3e. It is also worth noting that our estimated value of $H_{c2}(0)$ is well below the Pauli-Clogston limit.

We repeated the high-pressure experiments using different crystal flakes. Similar superconducting behavior with almost identical $T_c$ was observed. For comparison with 1T'-$MoTe_2$, we also measured $\rho(T)$ for the 2H-$MoTe_2$ single crystal at various pressure values. We found a pressure-induced metallization at 15 GPa (Supplementary Fig. 8), which is consistent with previous theoretical predictions[22]. However, in contrast, we did not detect any signature of superconductivity in the 2H phase for pressures up to 40 GPa.

**Discussion**

For $MoTe_2$, the superconducting behavior in the low-$P$ region clearly differs from that in the high-$P$ region. Under quite low $P$, the sharp increase in $T_c$ is concomitant with a strong suppression of the structural transition, which is reminiscent of observations for other superconductors with various kinds of competing phase transitions. The drastic increase of the $T_c$ occurs within the $T_d$ phase, which is shown by DFT calculations to be a Weyl semimetal (Supplementary Fig. 9a and Supplementary Note 2) with a band structure around the Fermi level which is extremely sensitive to changes in the lattice constants[24, 34]. Thus, one can expect that dramatic structural and electronic instabilities emerge in the low-*P* region, which may account for the strong enhancement of $T_c$. At higher pressures, the topologically trivial (due to inversion and time reversal symmetry) 1T' phase (Supplementary Fig. 9b and Supplementary Note 2) remains stable in the whole temperature range. Although within this phase $T_c$ still continues to increase up to its maximum value, the rate of the increase is significantly lower and this growth is naturally explained by the increase of the electronic density of states at the Fermi level in the 1T' phase (Supplementary Fig. 9c). Thorough exploration of superconductivity in $MoTe_2$ from both experimental and theoretical perspectives is required.

## Methods

### Single–crystal growth

1T'-$MoTe_2$ crystals were grown via chemical vapor transport using polycrystalline $MoTe_2$ powder and $TeCl_4$ as a transport additive[35]. Molar quantities of Mo (Sigma Aldrich 99.99%) were ground in combination with purified Te pieces (Alfa Aesar 99.99%), pressed into pellets, and heated in an evacuated quartz tube at 800 °C for 7 days. Crystals were obtained by sealing 1 g of this powder and $TeCl_4$ (3 mg/ml) in a quartz ampoule, which was then flushed with Ar, evacuated, sealed, and heated in a two-zone furnace. Crystallization was conducted from ($T_2$) 1000 to ($T_1$) 900 °C. The quartz ampoule was then quenched in ice water to yield the high-temperature monoclinic phase. The obtained crystals were silver-gray and rectangular in shape. 2H-$MoTe_2$ crystals were grown using a similar method, but without quenching.

**Structural and transport measurements at ambient pressure**

The structures of the $MoTe_2$ crystals were investigated using single-crystal x-ray diffraction (SXRD) with Mo $K_a$ radiation. In order to analyze the atomic structure of the material, high-angle annular dark-field scanning transmission electron microscopy (HAADF-STEM) was performed. The dependence of the electrical resistivity $\rho$ on temperature $T$ was measured using a conventional four-probe method (low-frequency alternating current, PPMS, Quantum Design). Temperatures down to 0.08 K were achieved using a home-built adiabatic demagnetization stage. The pulsed magnetic field experiments were conducted at the Dresden High Magnetic Field Laboratory (Helmholtz-Zentrum Dresden-Rossendorf, HLD-HZDR).

**Experimental details of high pressure measurements**

A non-magnetic diamond anvil cell was used for $\rho$ measurements under $P$ values of up to 40 GPa. A cubic BN/epoxy mixture was used for the insulating gaskets and Pt foil was employed in the electrical leads. The diameters of the flat working surface of the diamond anvil and the sample chamber were 500 and 200 μm, respectively. The initial sample thickness was ≈ 40 μm. The value of $\rho$ was measured using the dc current in van der Pauw technique in a customary cryogenic setup (lowest achievable temperature 1.5 K) at zero magnetic field, and the magnetic field measurements were performed on a PPMS. Pressure was measured using the ruby scale[36] by measuring the luminescence from small chips of ruby placed in contact with the sample.

Magnetization was measured on $MoTe_2$ ($m$ = 3.1 mg) in a pressure cell ($m$ = 170 mg) for $P \leq 0.7$ GPa and $T \geq 0.5$ K (Quantum Design MPMS, iQuantum $^3$He insert). Shielding (after zero-field cooling) and Meißner effect curves (in field-cooling) were recorded.

The high-$P$ Raman spectra were recorded using a customary micro-Raman spectrometer with a HeNe laser as the excitation source and a single-grating spectrograph with 1 cm$^{-1}$ resolution. Raman scattering was calibrated using Ne lines with an uncertainty of ±1 cm$^{-1}$.

High-pressure diffraction experiments have been performed at ID09A synchrotron beamline (ESRF) using monochromatic x-ray beam ($E$ = 30 keV, $\lambda$ = 0.413 Å) focused to 15 × 10 $\mu m^2$ on the sample[37]. We used a membrane-driven high-pressure cell equipped with Boehler-Almax seats and diamond anvil design, allowing an opening cone of 64°. The culet size was 600 $\mu$m and the sample was loaded together with He as pressure transmission medium into a hole in a stainless steel gasket preindented to about 80 $\mu$m with an initial diameter of 300 $\mu$m. Low temperature data were collected in a He-flow cryostat. Single-crystal data have been collected by a vertical-acting $\omega$-axis rotation, with an integrated step scan of 0.5° and a counting time of 1s per frame. Diffraction intensities have been recorded with a Mar555 flat-panel detector. Diffraction data have been processed and analyzed with CrysAlisPro-171.37.35 and Jana2006 software. Pressures were measured with the ruby fluorescence method [36].

**Density functional theory calculations**

Density-functional theory (DFT) calculations were performed using the Vienna Ab-initio Simulation Package (VASP) with projected augmented wave (PAW) potential[38,39]. The exchange and correlation energy was considered at the generalized gradient approximation (GGA) level for the geometry optimization[40], and the electronic structure was calculated using the hybrid functional (HSE06)[41]. Spin-orbital coupling was included in all calculations. Van der Waals corrections were included via a pair-wise force field of the Grimme method [42]. In the lattice relaxation, the volumes were fixed while lattice constants and atomic positions were optimized. The pressure was derived by fitting the total energy dependence on the volume with the Murnaghan equation[43]. After checking the *k* convergence, the 24×12×8 and 7×5×3 *k*-meshes with Gaussian-type smearing were used for the GGA (Supplementary Fig. 10) and HSE06 calculations, respectively. The band structures, density of states and Fermi surfaces were interpolated in a dense *k*-mesh of 200×200×200 using the maximally localized Wannier functions[44] extracted from HSE06 calculations.

**Acknowledgments**

Y. Qi acknowledges financial support from the Alexander von Humboldt Foundation. We would like to thank C. Klausnitzer & M. Nicklas for their help with high-pressure magnetic measurements. This work was financially supported by the Deutsche Forschungsgemeinschaft (DFG, Project No. EB 518/1-1 of DFG-SPP 1666 "Topological Insulators") and by a European Research Council (ERC) Advanced Grant, No. (291472) "Idea Heusler".

**Author Contributions.**

M. A., C. R. and V. S. prepared the samples and performed XRD structural characterization, E. P., P. W. and R. H. performed TEM studies, W. S. performed ambient pressure transport measurements and Meißner effect measurements at low pressures, C. S., T. F. and E. K. performed magneto-transport measurements, Y. Q., P. N., O. B., and S. M. performed high pressure electrical resistivity, Raman spectroscopy and magnetic susceptibility measurements, M. H. performed high pressure single crystal XRD studies, S. W., Y. S. and B. Y. carried out the theoretical calculations. All authors discussed the results of the studies. Y. Q., B. Y., W. S. and S. M. co-wrote the paper. All authors commented on the manuscript.

**Competing financial interests.**

The authors declare no competing financial interests.

## References

1. Wilson, J. A. & Yoffe, A. D. The Transition Metal Dichalcogenides Discussion and interpretation of the observed optical, electrical and structural properties. *Adv. Phys.* **18**, 193-335 (1969).
2. Klemm, R. A. Pristine and intercalated transition metal dichalcogenide superconductors. *Physica C* **514,** 86–94 (2015).
3. Morris, R. C., Coleman, R. V. & Bhandari, R. Superconductivity and magnetoresistance in $NbSe_2$. *Phys. Rev. B* **5,** 895–901 (1972).
4. Morosan, E. *et al*. Superconductivity in $Cu_xTiSe_2$. *Nature Phys.* **2**, 544-550 (2006).
5. Moncton, D. E., Axe, J. D. & DiSalvo, F. J. Neutron scattering study of the charge-density wave transitions in 2H-$TaSe_2$ and 2H-$NbSe_2$. *Phys. Rev. B* **16**, 801–819 (1977).
6. Qian, X., Liu, J., Fu, L. & Li, J. Quantum spin Hall effect in two-dimensional transition metal dichalcogenides. *Science* **346**, 1344-1347 (2014).
7. Xu, X., Yao, W., Xiao D. & Heinz T. F. Spin and pseudospins in layered transition metal dichalcogenides. *Nature Phys.* **10**, 343–350 (2014).
8. Bates, J. B., Gruzalski, G. R., Dudney, N. J., Luck, C. F. & Yu, X. Rechargeable thin-film lithium batteries. *Solid State Ionics* **70/71**, 619–628 (1994).

9. Li, Y., Wang, H., Xie. L., Liang, Y., Hong, G. & Dai, H. $MoS_2$ nanoparticles grown on graphene: an advanced catalyst for the hydrogen evolution reaction. *J. Am. Chem. Soc.* **133**, 7296–7299 (2011).

10. Zhang, Y. J., Oka, T., Suzuki, R., Ye, J. T. & Iwasa, Y. Electrically switchable chiral light-emitting transistor. *Science* **344**, 725-728 (2014).

11. Lin, Y.-C., Dumcenco, D. O., Huang, Y.-S. & Suenaga, K. Atomic mechanism of the semiconducting-to-metallic phase transition in single-layered $MoS_2$. *Nature Nanotech.* **9**, 391-396 (2014).

12. Clarke, R., Marseglia, E. & Hughes, H. P. A low-temperature structural phase transition in β-$MoTe_2$. *Philos. Mag. B* **38**, 121-126 (1978).

13. Puotinen, D. & Newnhan, R. E. The crystal structure of $MoTe_2$. *Acta Crystallogr.* **14**, 691-692 (1961).

14. Zandt, T., Dwelk, H., Janowitz, C. & Manzke, R. Quadratic temperature dependence up to 50 K of the resistivity of metallic $MoTe_2$. *J. Alloy. Compd.* **442**, 216-218 (2007).

15. Brown, B. E. The crystal structures of $WTe_2$ and high-temperature $MoTe_2$. *Acta Crystallogr.* **20**, 268–274 (1966).

16. Ali, M. N. *et al.* Large, non-saturating magnetoresistance in $WTe_2$. *Nature* **514**,

205-208 (2014).

17. Ali, M. N. *et al*. Correlation of crystal quality and extreme magnetoresistance of $WTe_2$. *Europhys*. *Lett*. **110**, 67002 (2015).

18. Pan, X.-C. *et al*. Pressure-driven dome-shaped superconductivity and electronic structural evolution in tungsten ditelluride. *Nature Commun*. **6**, 7805 (2015).

19. Kang, D. *et al*. Superconductivity emerging from suppressed large magnetoresistant state in $WTe_2$. *Nature Commun*. **6**, 7804 (2015).

20. Zhu, Z. *et al*. Quantum Oscillations, Thermoelectric Coefficients, and the Fermi Surface of Semimetallic $WTe_2$. *Phys. Rev. Lett*. **114**, 176601 (2015).

21. Soluyanov, A. *et al*. Type II Weyl Semimetals. Preprint at http://arxiv.org/abs/1507.01603 (2015).

22. Rifliková, M., Martoňák, R. & Tosatti, E. Pressure-induced gap closing and metallization of $MoSe_2$ and $MoTe_2$. *Phys. Rev. B* **90**, 035108 (2014).

23. Fu, L. & Kane, C. L. Superconducting Proximity Effect and Majorana Fermions at the Surface of a Topological Insulator. *Phys. Rev. Lett.* **100**, 096407 (2008).

24. Sun, Y., Wu, S.-C., Ali, M. N., Felser, C. & Yan, B. Prediction of the Weyl semimetal in the orthorhombic $MoTe_2$. *Phys. Rev. B* **92**, 161107 (2015).

25. Cho, G. Y., Bardarson, J. H., Y.-M. Lu, & Moore, J. E. Superconductivity of doped

Weyl semimetals: Finite-momentum pairing and electronic analog of the $^3$He-A phase. *Phys. Rev. B* **86**, 214514 (2012).

26. Wei, H., Chao, S.-P. & Aji, V. Odd-parity superconductivity in Weyl semimetals. *Phys. Rev. B* **89**, 014506 (2014).
27. Hosur, P., Dai, X., Fang, Z. & Qi, X.-L. Time-reversal-invariant topological superconductivity in doped Weyl semimetals. *Phys. Rev. B* **90**, 045130 (2014).
28. Jian, S.-K., Jiang, Y.-F. & Yao, H. Emergent Spacetime Supersymmetry in 3D Weyl Semimetals and 2D Dirac Semimetals. *Phys. Rev. Lett.* **114**, 237001 (2015).
29. Keum, D. H. *et al.* Bandgap opening in few-layered monoclinic $MoTe_2$. *Nature Phys.* **11**, 482–487 (2015).
30. Hughes, H. P. & Friend, R. H. Electrical resistivity anomaly in $\beta$–$MoTe_2$. *J. Phys. C: Solid State Phys.* **11,** L103-L105 (1978).
31. Hulliger, F. Crystal Chemistry of the Chalcogenides and Pnictides of the Transition Elements, in *Structure and Bonding 83-229*, (Structure and Bonding vol. 4, Springer-Verlag, New York, Heidelberg, Berlin 1968).
32. Suderow, H., Tissen, V. G., Brison, J. P., Martínez, J. L. & Vieira, S. Pressure induced effects on the Fermi surface of superconducting 2H-$NbSe_2$. *Phys. Rev. Lett.* **95**, 117006 (2005).

33. Müller, K. H., Fuchs, G., Handstein, A., Nenkov, K., Narozhnyi, V.N., & Eckert, D. The upper critical field in superconducting $MgB_2$. *J. Alloy. Compd.* **322**, L10-L13 (2001).

34. Wang, Z. *et al.* $MoTe_2$: Weyl and Line Node Topological Metal. Preprint at http://arxiv.org/abs/1511.07440 (2015).

35. Fourcaudot, G., Gourmala, M., & Mercier, J. Vapor phase transport and crystal growth of molybdenum trioxide and molybdenum ditelluride. *J. Cryst. Growth* **46**, 132-135 (1979).

36. Mao, H. K., Xu, J., & Bell, P. M. Calibration of the ruby pressure gauge to 800 kbar under quasi-hydrostatic conditions. *J. Geophys. Res.* **91**, 4673-4676 (1986).

37. Merlini, M. & Hanfland, M. Single-crystal diffraction at megabar conditions by synchrotron radiation. *High Pressure Res.* **33**, 511-522 (2013).

38. Kresse, G. & Hafner, J. *Ab initio* molecular dynamics for open-shell transition metals. *Phys. Rev. B* **48**, 13115-13118 (1993).

39. Kresse, G. & Furthmüller, J. Efficiency of ab-initio total energy calculations for metals and semiconductors using a plane-wave basis set. *Comp. Mater. Sci.* **6**, 15-50 (1996).

40. Perdew, J. P., Burke, K., & Ernzerhof, M. Generalized Gradient Approximation

Made Simple. *Phys. Rev. Lett.* **77**, 3865-3868 (1996).

41. Heyd, J., Scuseria, G. E. & Ernzerhof, M. Hybrid functionals based on a screened Coulomb potential. *J. Chem. Phys.* **118**, 8207-8215 (2003).

42. Grimme, S. Semiempirical GGA-Type Density Functional Constructed with a Long-Range Dispersion Correction. *J. Comput. Chem.* **27**, 1787-1799 (2006).

43. Murnaghan, F. D. The Compressibility of Media under Extreme Pressures. *P. Natl. Acad. Sci. U.S.A.* **30**, 244-247 (1944).

44. Marzari, N. & Vanderbilt, D. Maximally localized generalized Wannier functions for composite energy bands. *Phys. Rev. B* **56**, 12847-12865 (1997).

**Figure Captions**

**Figure 1. $MoTe_2$ crystal structure.** (**a**) HAADF-STEM image of 1T'-$MoTe_2$ along the [100] zone (scale bar, 0.5 nm). The red rectangle shows HAADF simulated image, and the red and blue spheres in the yellow rectangle represent Te and Mo atoms, respectively. (**b**) Corresponding electron diffraction images. (**c**) 1T' and $T_d$-$MoTe_2$ crystal structures. (**d**) Energy-volume dependence for 1T' and $T_d$ phases from DFT calculations.

**Figure 2. Resistivity of 1T'-$MoTe_2$ at ambient pressure.** (**a**) Temperature-dependent resistivity at near zero pressure. Inset: Anomaly with hysteresis observed at approximately 250 K. This hysteresis is associated with the structural phase transition from 1T'-$MoTe_2$ to $T_d$-$MoTe_2$. (**b)** Resistivity detail from 0.08−1.2 K. Superconductivity is observed with onset at ≈ 0.3 K and zero resistance at $T_c$ = 0.10 K.

**Figure 3. Transport properties of 1T'-$MoTe_2$ as a function of pressure. (a)** Electrical resistivity as a function of temperature for pressures of 0.76−34.9 GPa. The anomaly associated with the structural transition is completely suppressed with increasing pressure. (**b**), (**c**) Electrical resistivity as a function of temperature for pressures of 0.7−11.7 and 11.7−34.9 GPa, respectively. Clear electrical resistivity drops

and zero-resistance behavior are apparent. $T_c$ increases under increasing pressure and a dome-shaped superconducting phase in pressure-temperature space is observed for the maximum superconducting transition temperature corresponding to $T_c$ = 8.2 K at 11.7 GPa. (**d**) Temperature dependence of resistivity under different magnetic fields of up to 3 T at 11.2 GPa. (**e**) Temperature dependence of $MoTe_2$ upper critical field $H_{c2}$ determined using 90% points on resistivity transition curves. The red curve is the best-fit line.

**Figure 4. High pressure Raman spectroscopy and structural studies of 1T'-$MoTe_2$.** (**a**)**,** Pressure-dependent Raman signals for 1T'-$MoTe_2$ at room temperature. The Raman spectra contain two characteristic peaks due to the $A_g$ and $B_g$ vibrational modes of the 1T'-$MoTe_2$ structure. (**b**) Frequencies of $A_g$ and $B_g$ modes as function of pressure. The frequencies of both vibrational modes increase gradually and continuously as the pressure increases. (**c**) Pressure dependence of the monoclinic angle $\beta$ obtained from single crystal x-ray diffraction studies. Isothermal compression at room temperature (red circles) shows increase of the monoclinic distortion with pressure whereas reversible orthorhombic $T_d$ – monoclinic 1T' transition is observed in isothermal compression (filled blue circles) / decompression (open blue circles) run at 135 K. The error bars in (b) and (c) due to s. d. are smaller than the symbols size.

**Figure 5. $MoTe_2$ electronic phase diagram.** The black and green square represents the structural phase transition temperature $T_s$ obtained from resistivity and single crystal synchrotron x-ray diffraction data. The red, blue, and olive circles represent the $T_c$ extracted from various electrical resistance measurements, and the pink triangles represent the $T_c$ determined from the magnetization measurements. The error bars in due to s. d. are smaller than the symbols size.

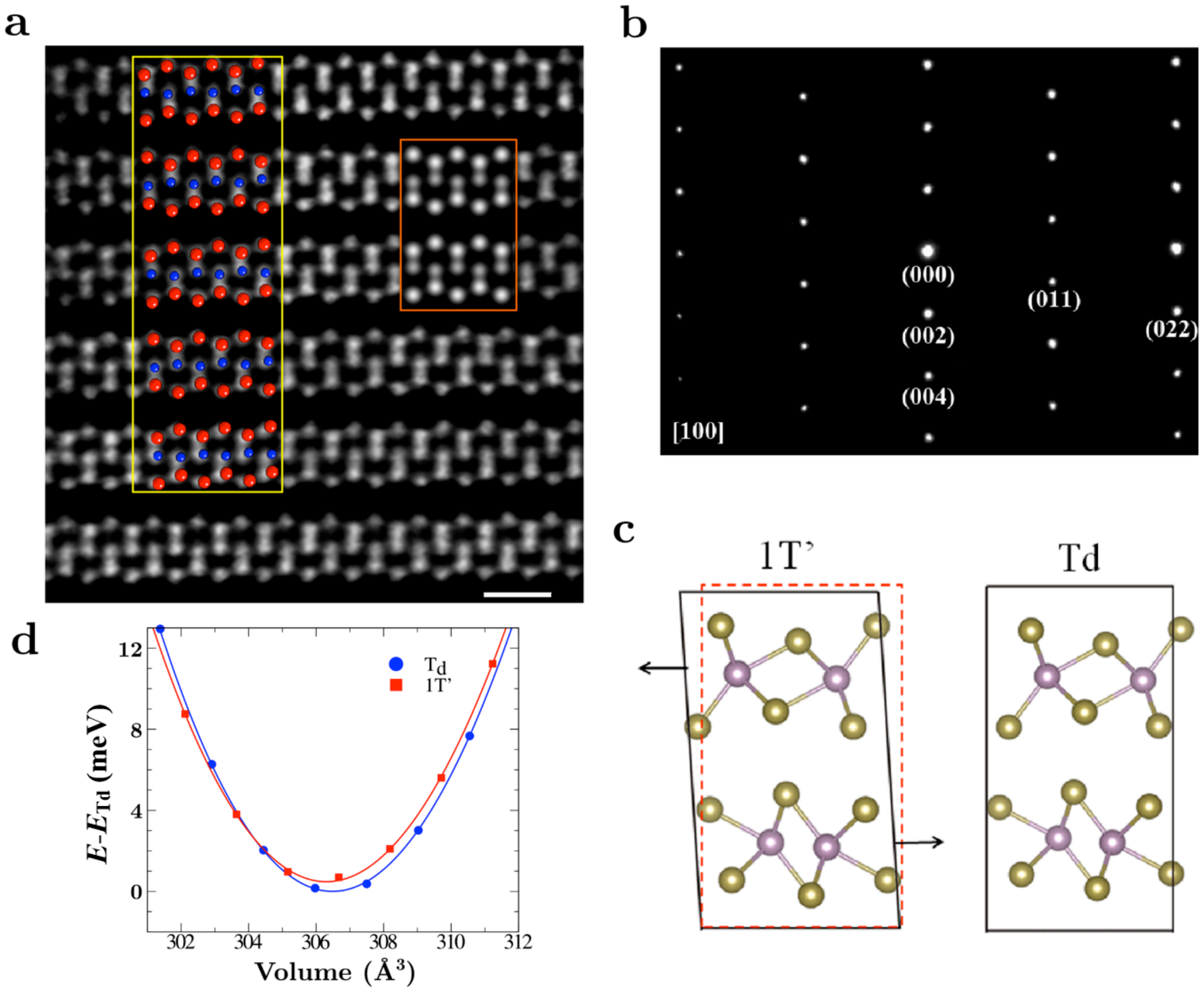

Fig. 1 Qi et al

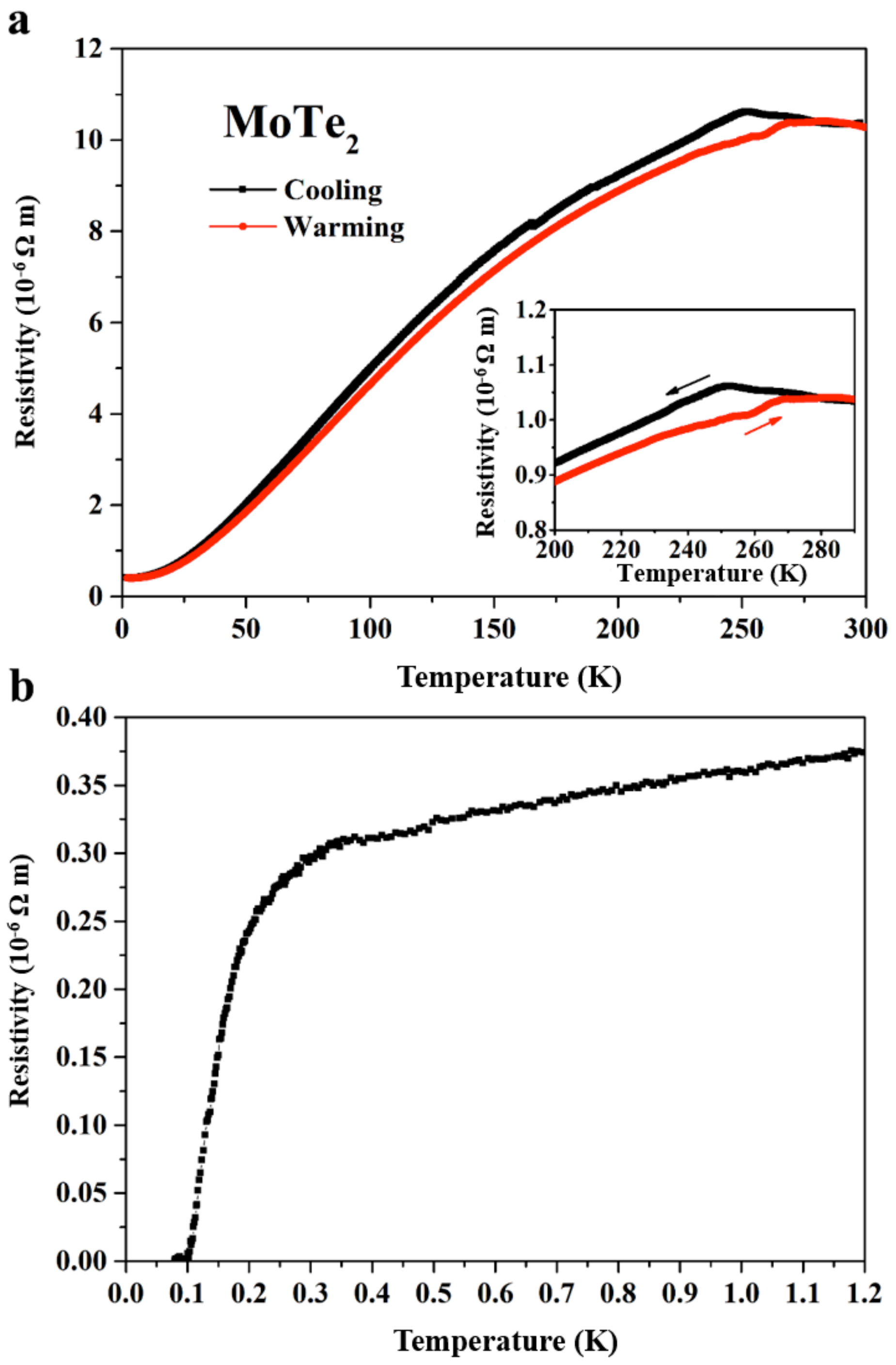
a
MoTe2
Cooling
Warming
Resistivity (10-6 Ω m)
Temperature (K)
b
Resistivity (10-6 Ω m)
Temperature (K)

Fig. 2 Qi et al.

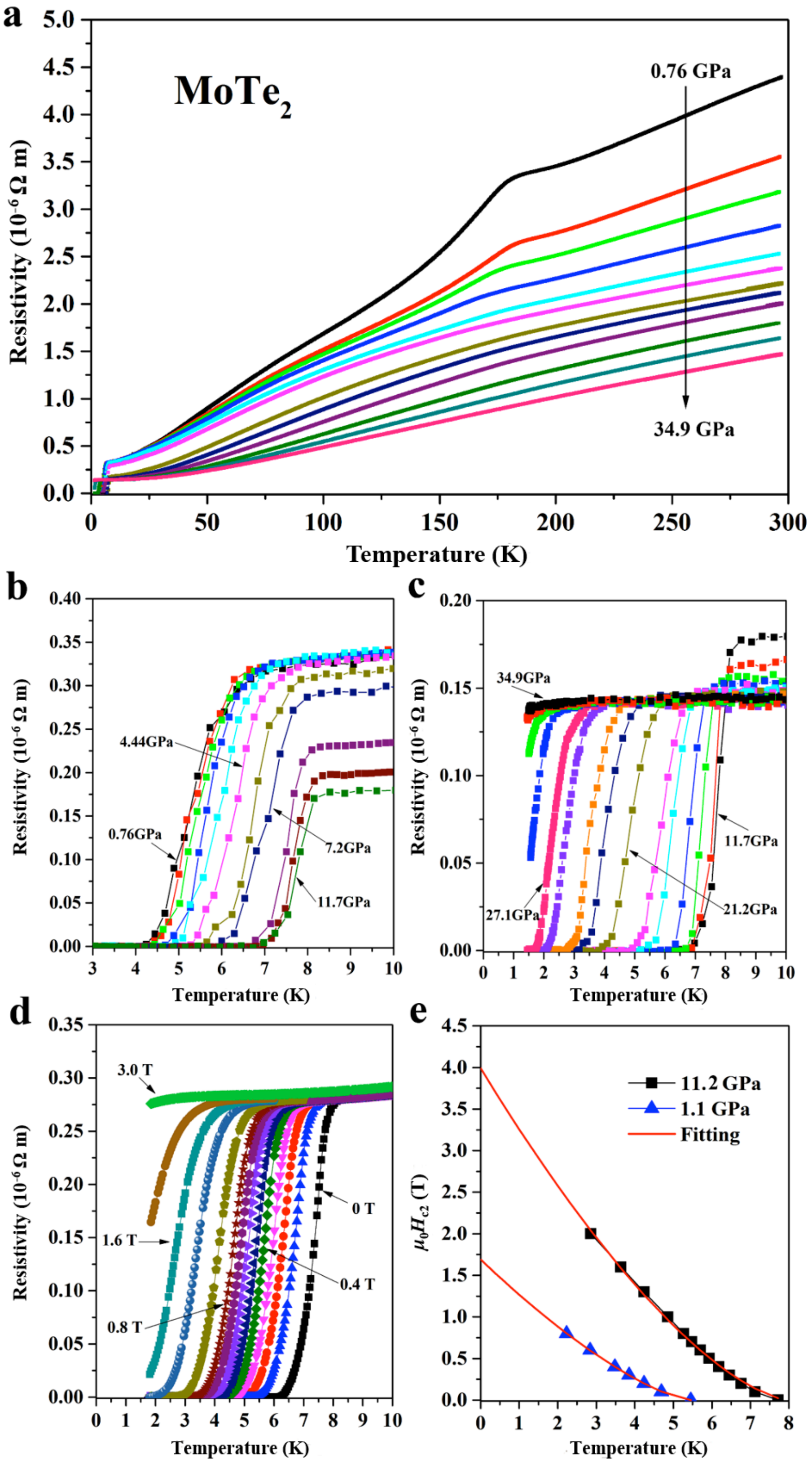

Fig. 3 Qi et al.

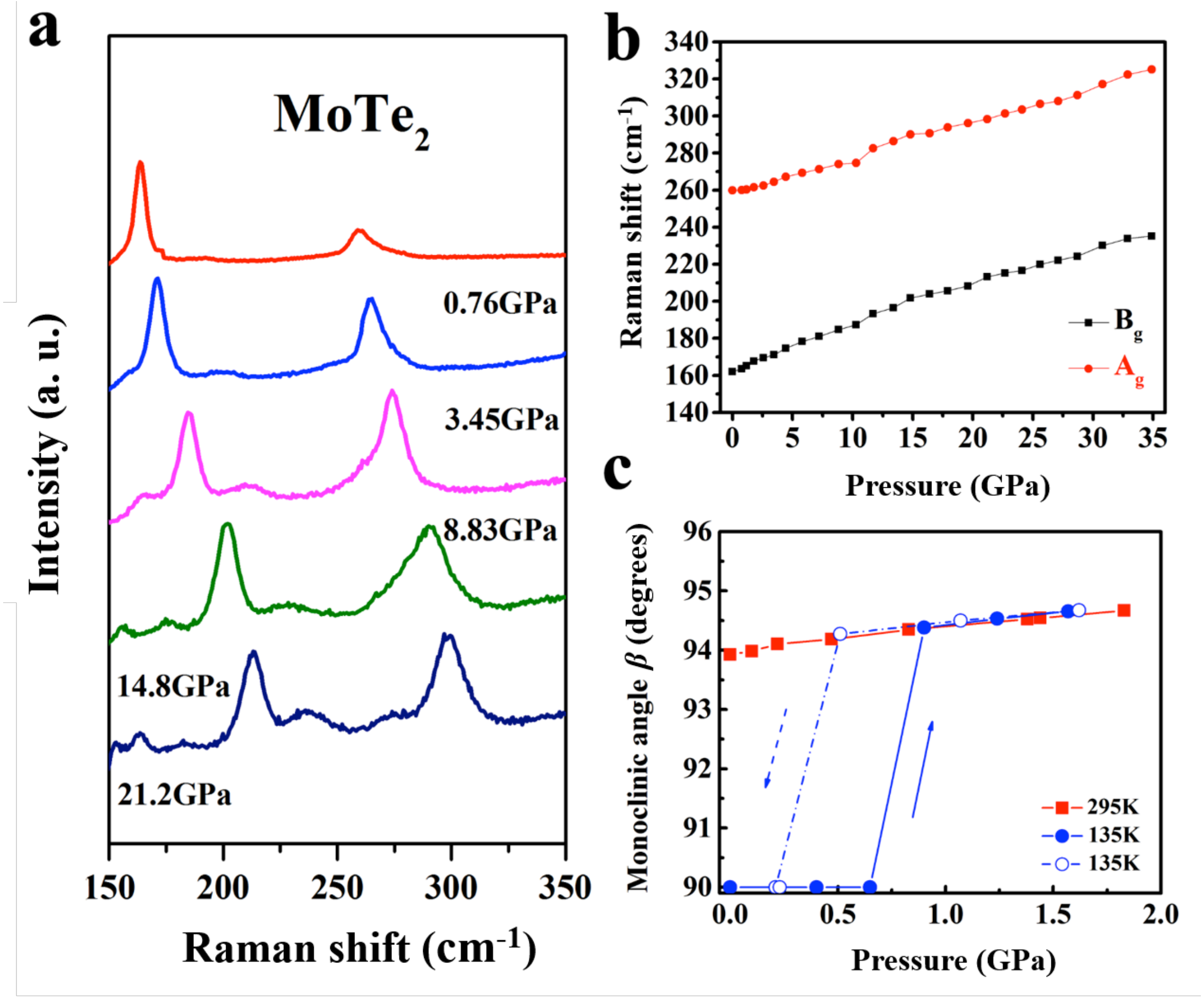

Fig. 4 Qi et al

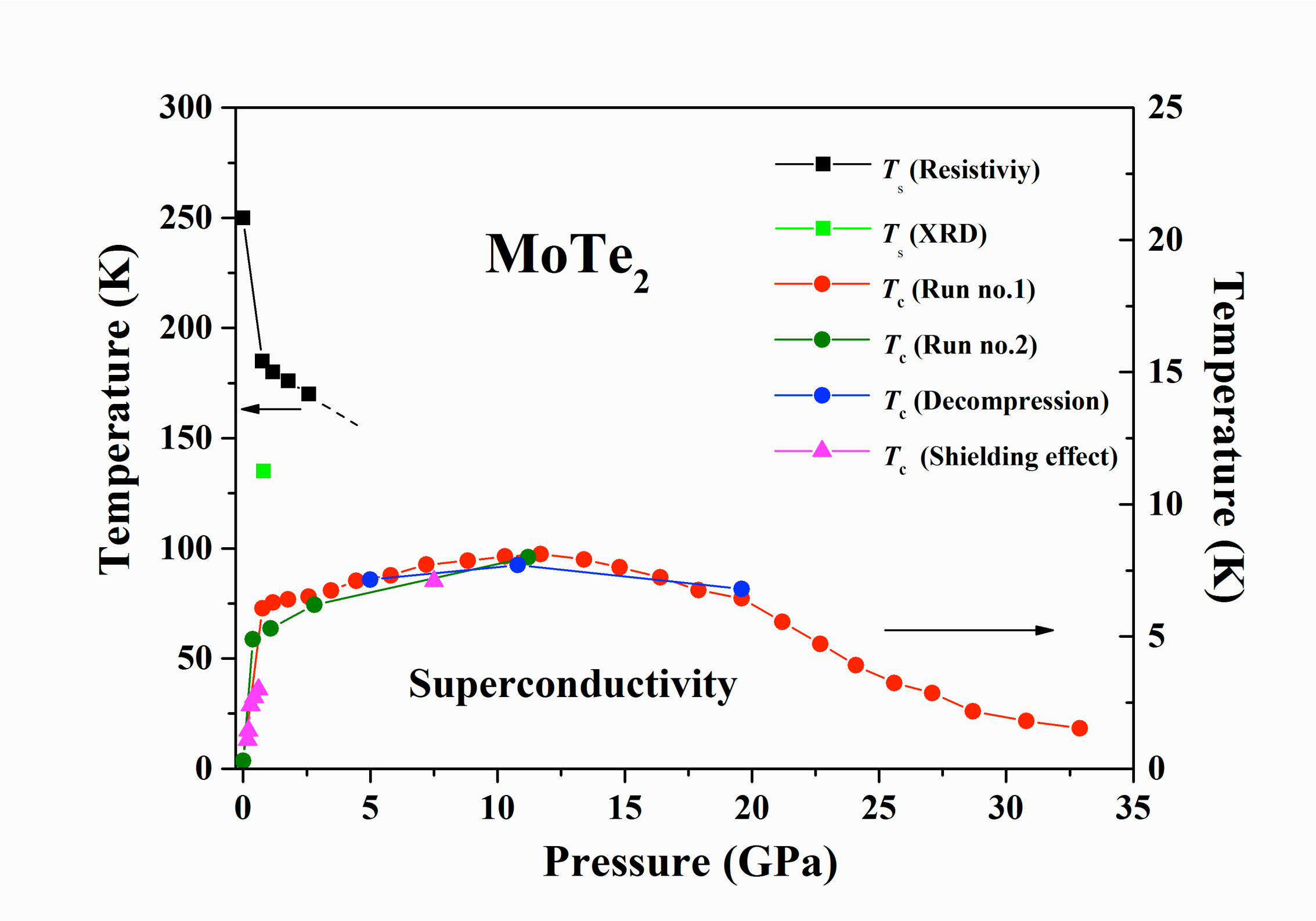

Fig. 5 Qi et al.

## Supplementary Figures

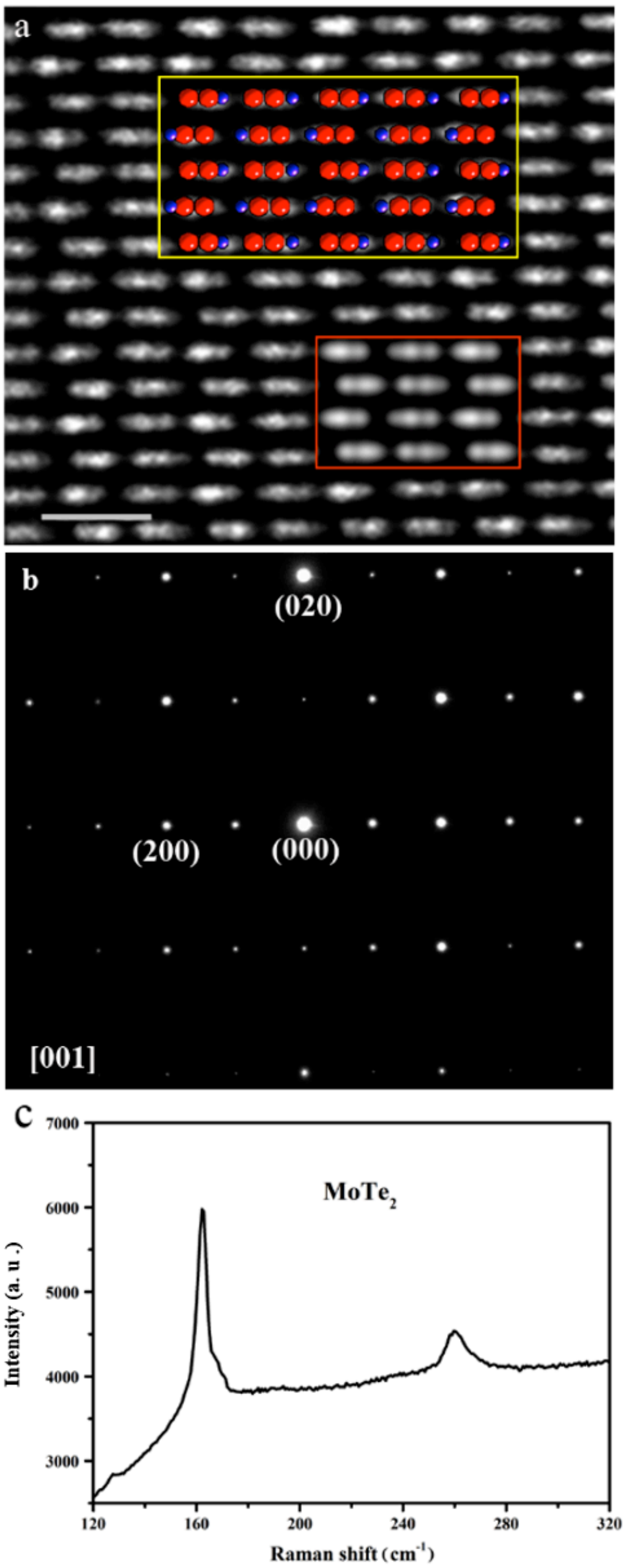

**Supplementary Figure 1. Crystal structure of 1T'-MoTe$_2$.** (**a**) HAADF-STEM image of 1T'-MoTe$_2$, looking down the [001] zone (scale bar, 0.5 nm). The area indicated by the red rectangle shows HAADF simulated image. The red spheres represent Te atoms and blue spheres represent Mo atoms. (**b**) The corresponding electron diffraction image. **c,** Raman signals for 1T'-MoTe$_2$ at ambient pressure.

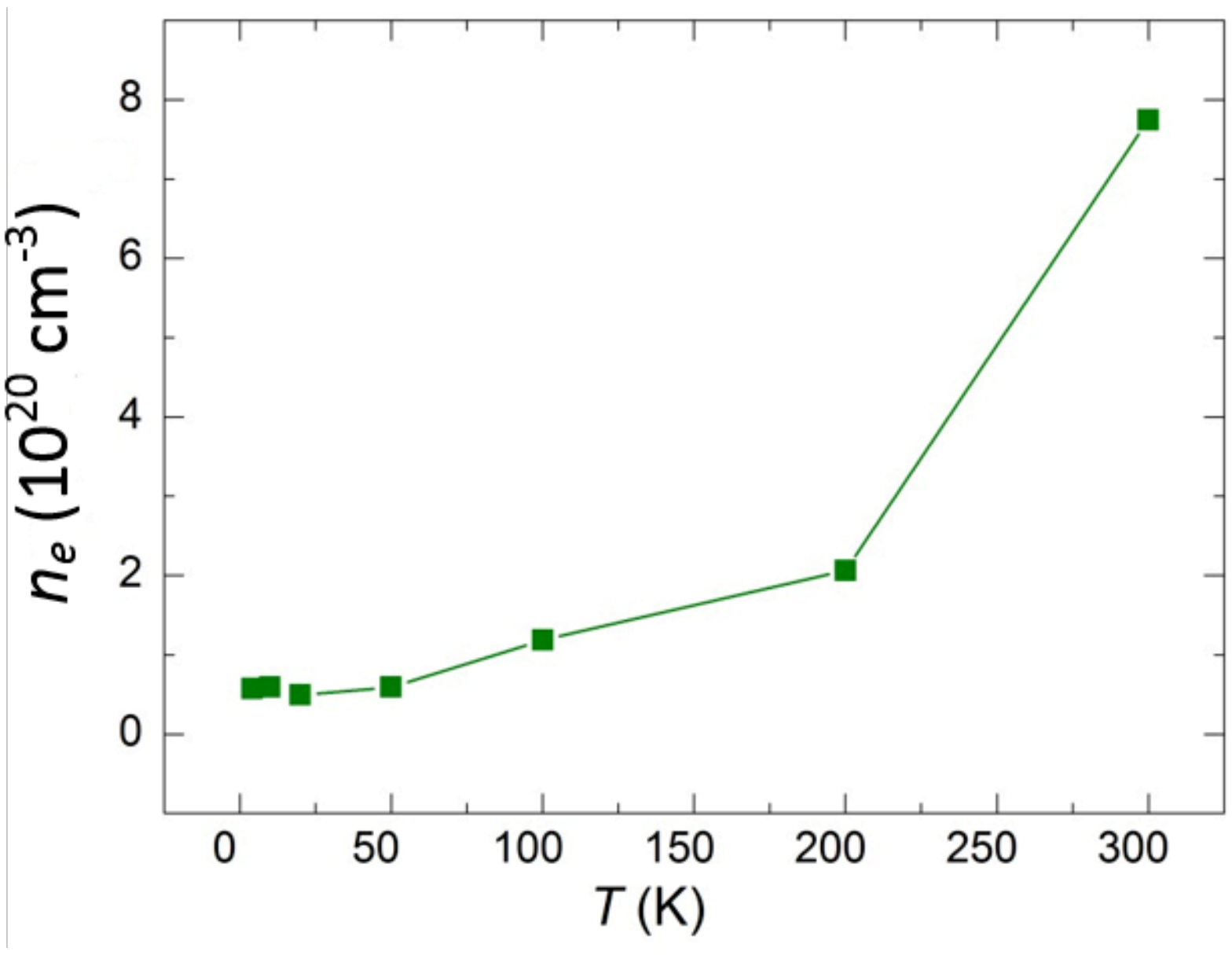

**Supplementary Figure 2. Electron charge carrier density of 1T'-$MoTe_2$.**

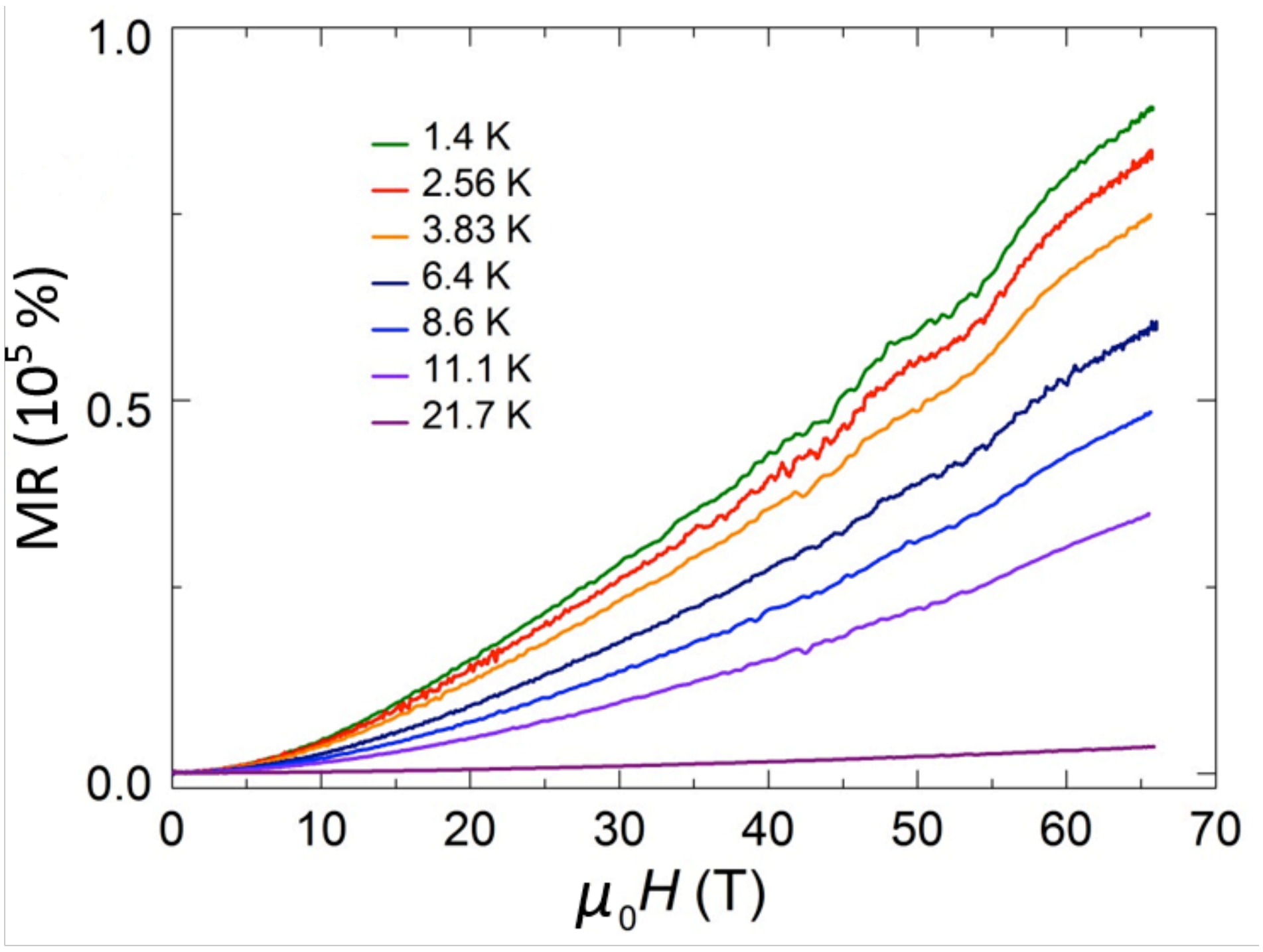

**Supplementary Figure 3. Magnetoresistance measured up to a field $\mu_0 H$ = 66 T.** Magneto-resistivity superimposed with quantum oscillations known as Shubnikov-de Haas (SdH) oscillations.

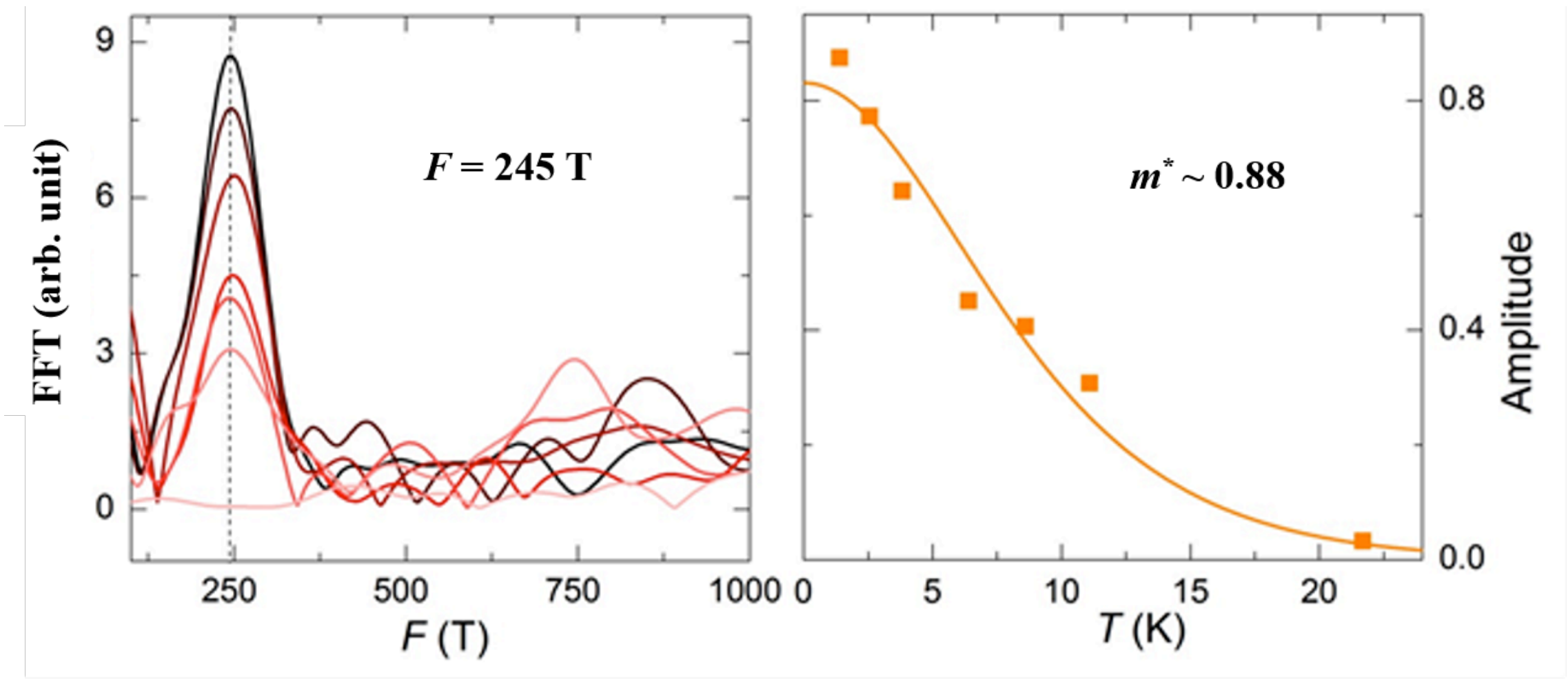

**Supplementary Figure 4. Fast Fourier transformation of SdH oscillations.** (**a**) The fast Fourier transformation (FFT) of the SdH oscillations. (**b**) Temperature dependent SdH oscillations amplitude.

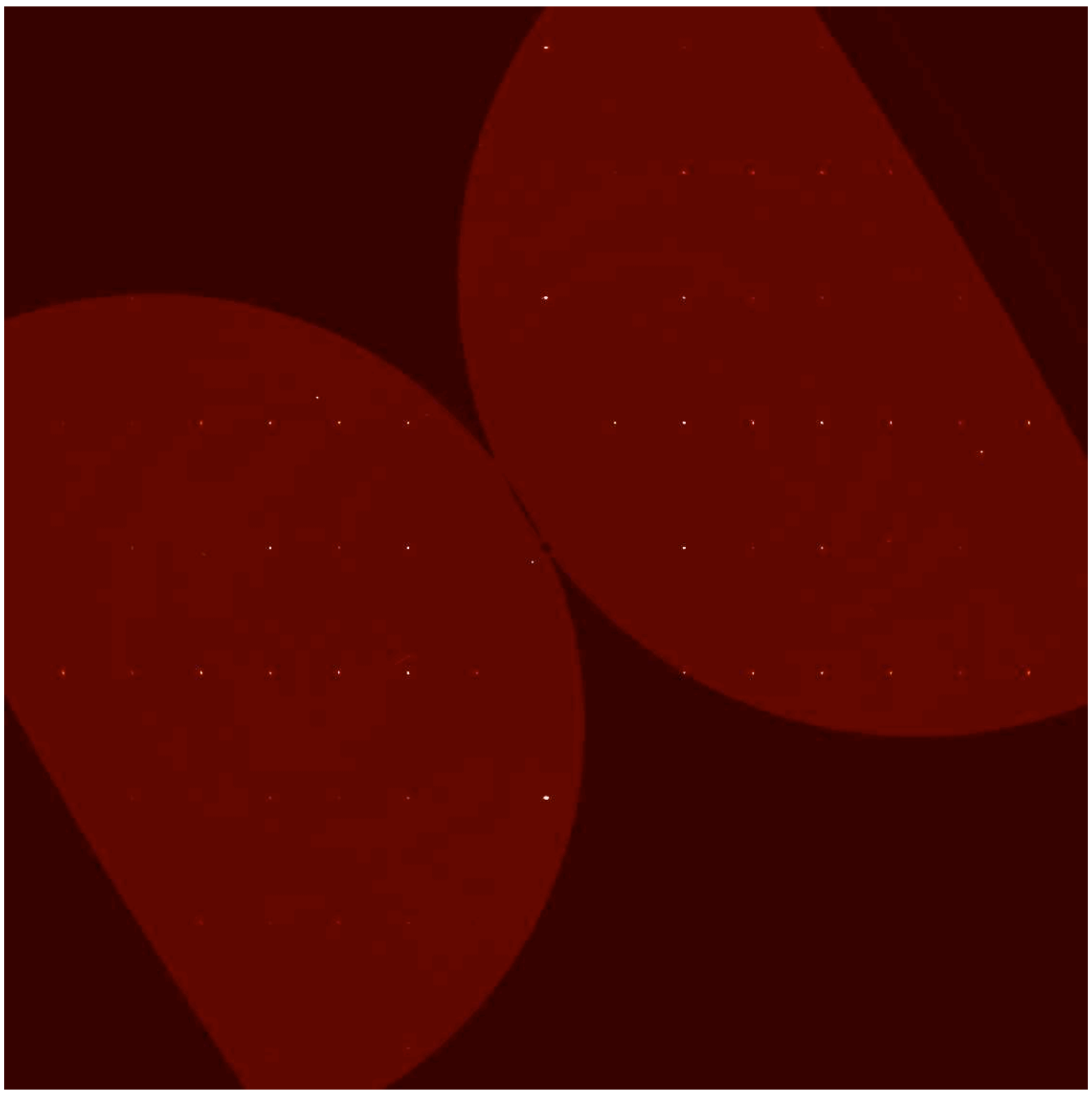

**Supplementary Figure 5. Reciprocal space reconstruction of the *hk*0 plane based on single-crystal diffraction data of *β*-$MoTe_2$ collected at pressure 1.56 GPa and temperature 120 K.**

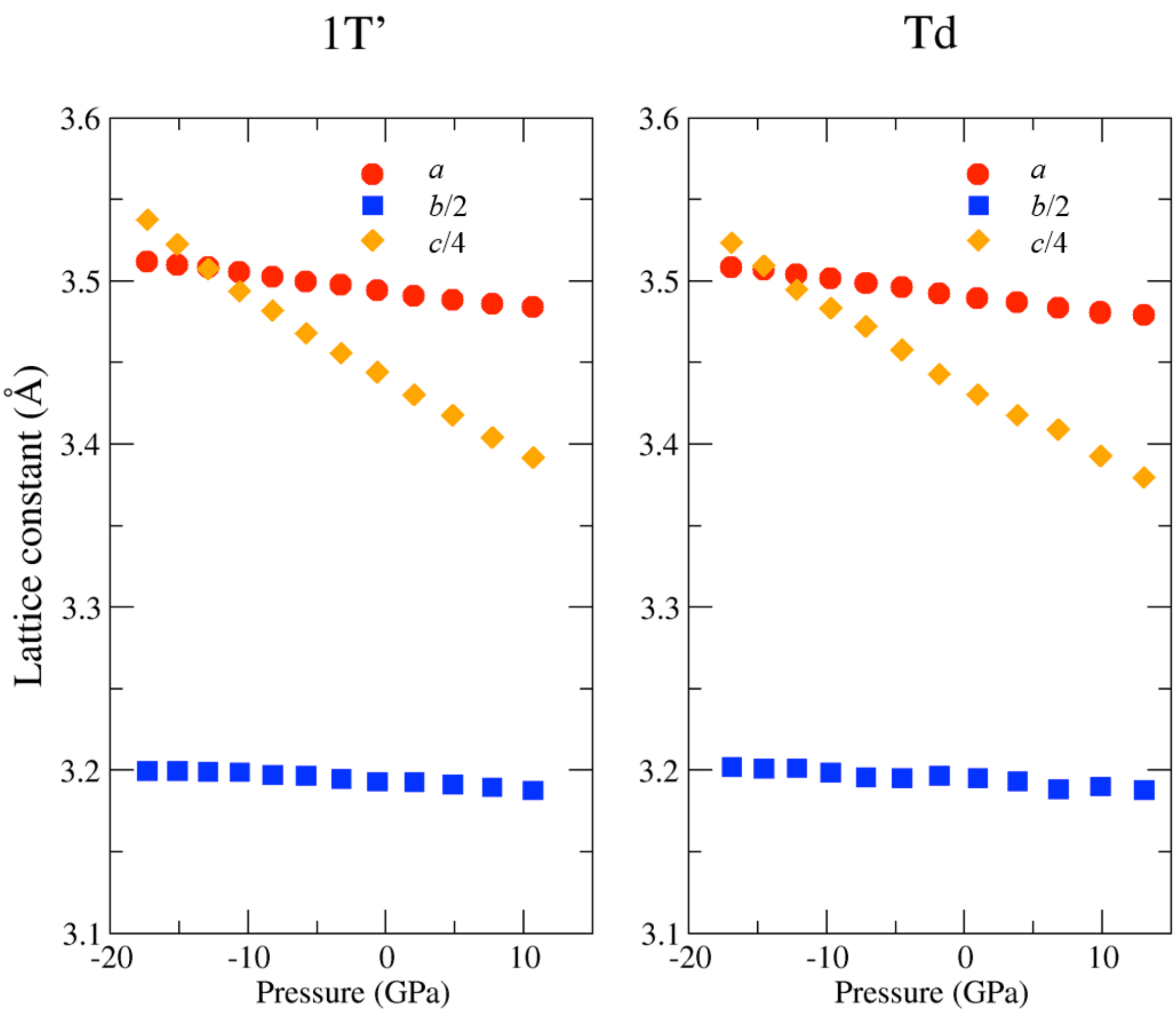

**Supplementary Figure 6. Lattice parameters as a function of pressure for $T_d$ and 1T' phases from GGA calculations.**

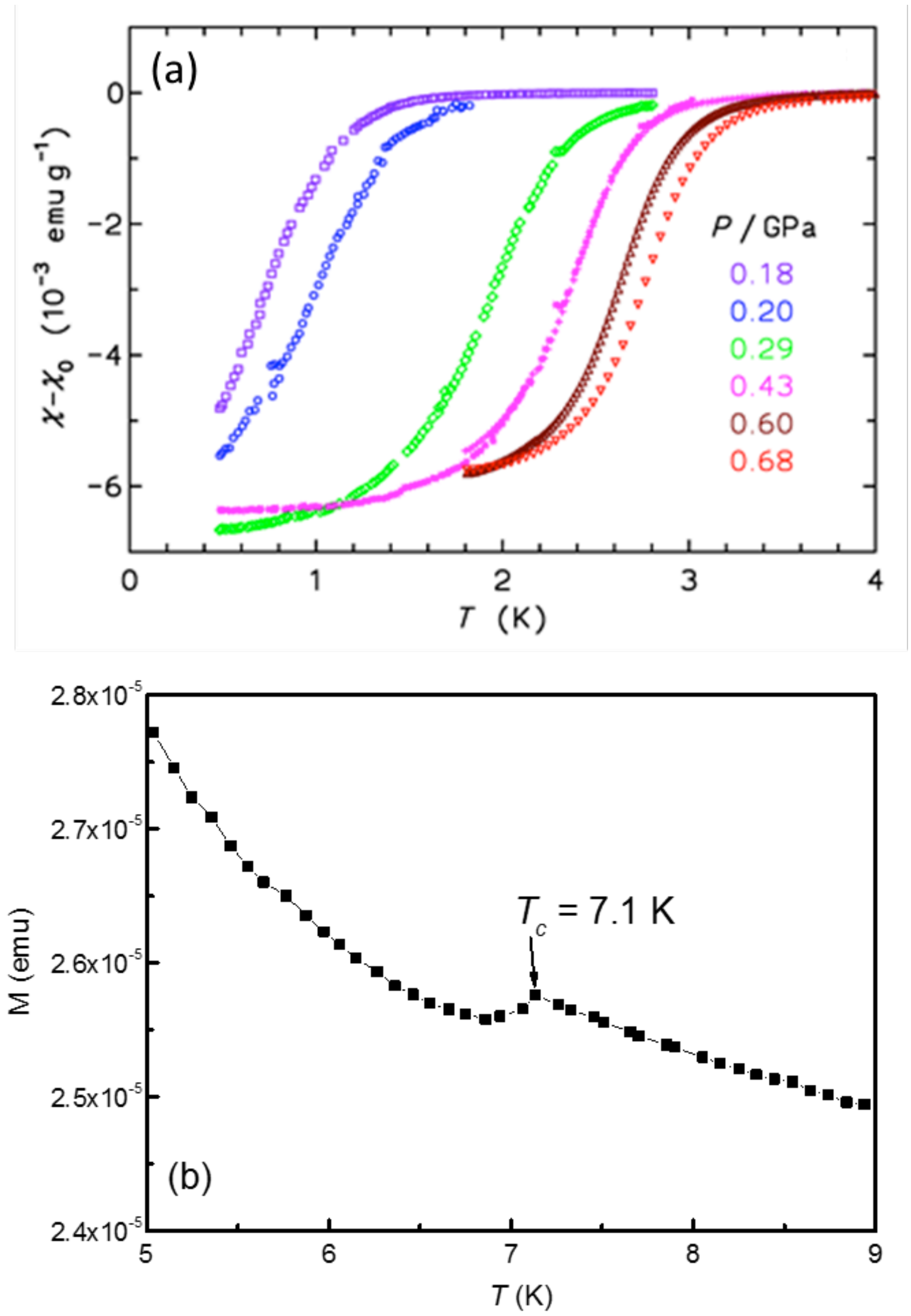

**Supplementary Figure 7. Magnetic shielding effect measurements at high pressure.** (**a**) The magnetic susceptibility for $P < 0.7$ GPa, (**b**) The raw magnetization signal at $P = 7.5$ GPa for $MoTe_2$, both as a function of temperature in a miniature high-pressure cell[1].

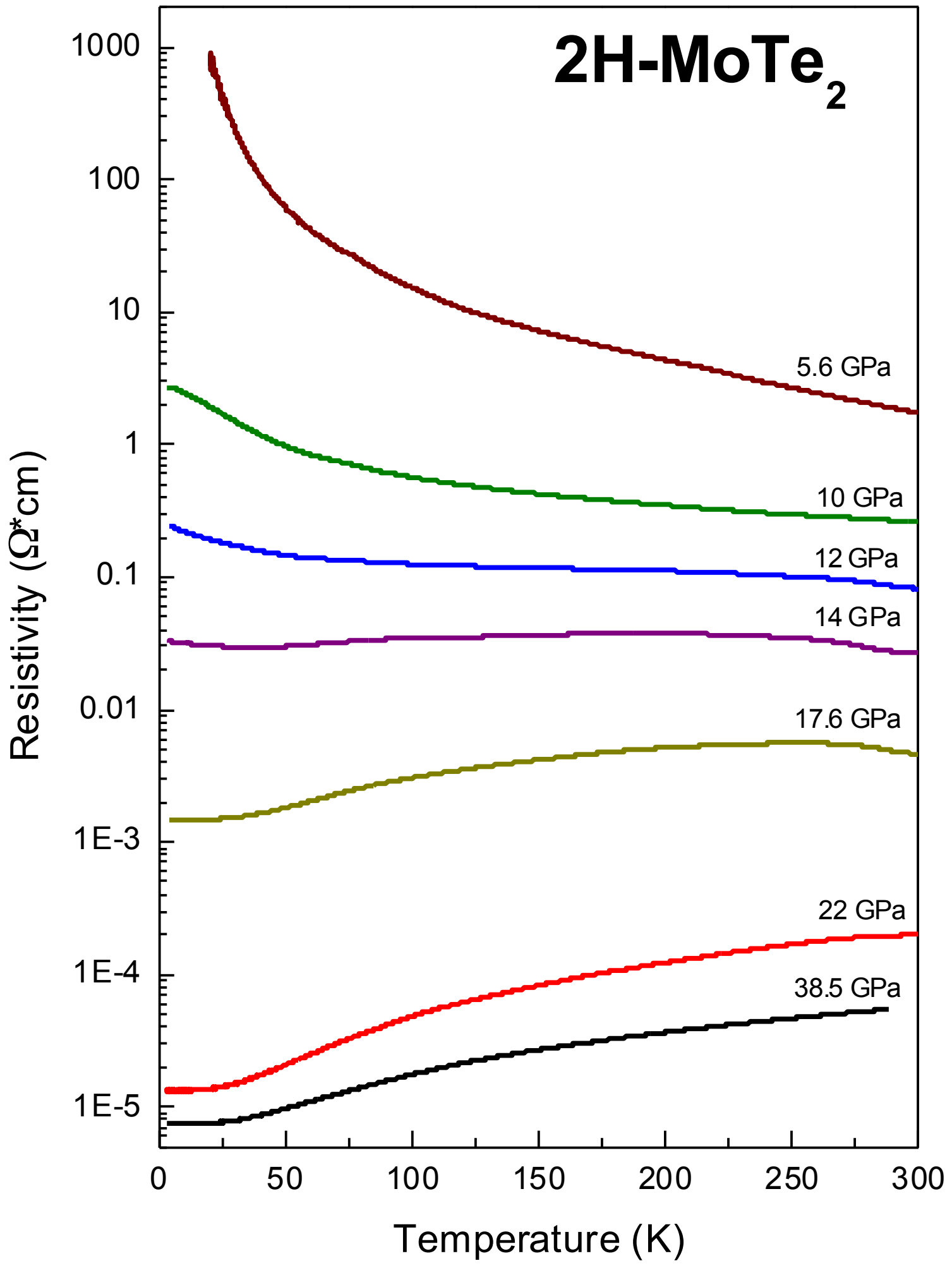

**Supplementary Figure 8. Temperature dependence of the resistivity of 2H-MoTe$_2$ under high pressure up to nearly 40 GPa.** At pressures below 15 GPa, 2H-MoTe$_2$ remains semiconducting. By 16 GPa, 2H-MoTe$_2$ shows metallic-like behavior the whole temperature range. Thus, 2H-MoTe$_2$ undergoes an insulator-to-metal transition at around 16 GPa, in good agreement with recent theoretical predictions. No signs of superconductivity ($T_c$ > 1.5 K) have been found up to nearly 40 GPa.

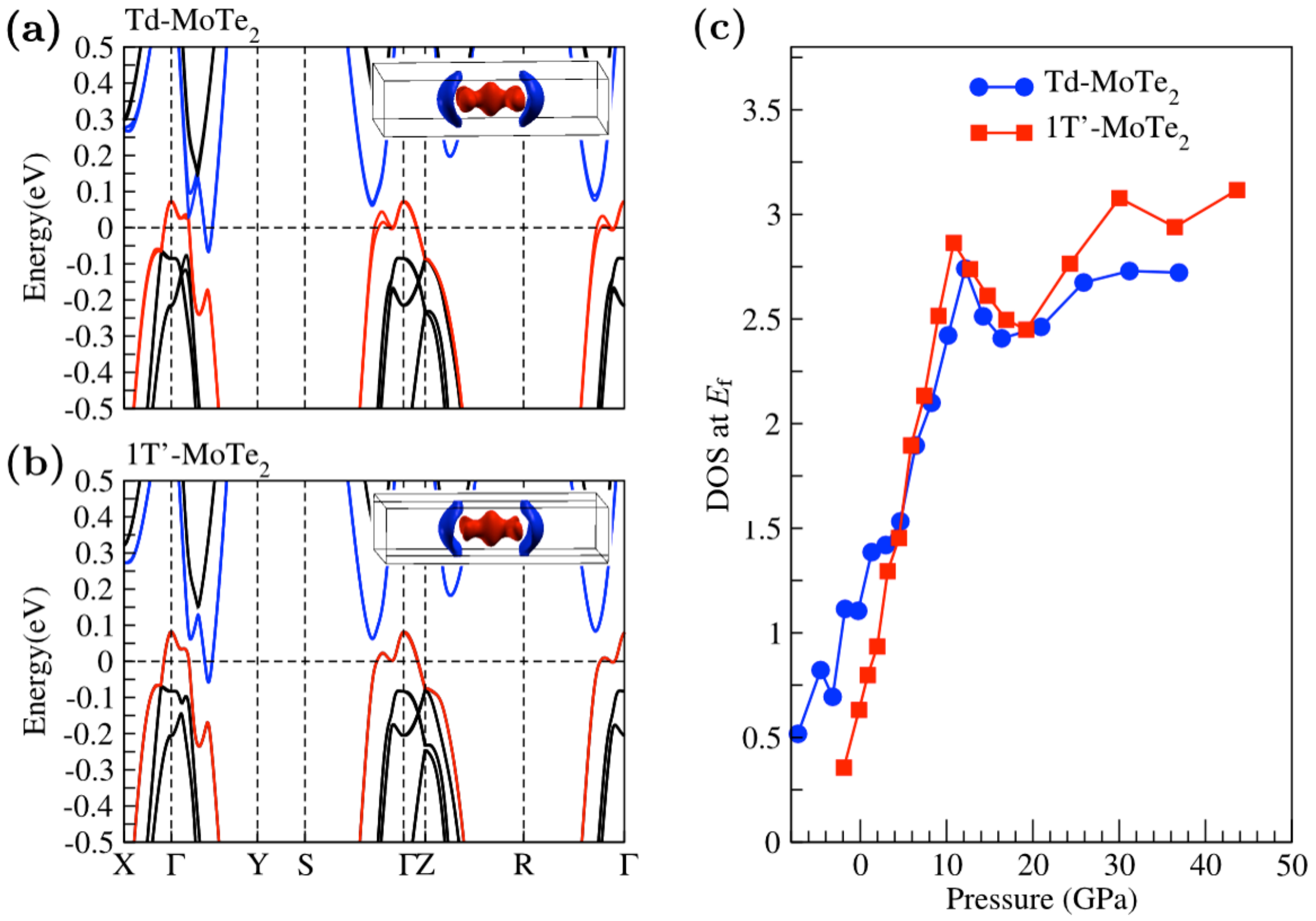

**Supplementary Figure 9. Electronic structures obtained from DFT calculations.** (**a**) and (**b**) Electronic band structures and Fermi surfaces for $MoTe_2$ in $T_d$ and 1T'-$MoTe_2$ phases respectively with experimental lattice constants. Red and blue Fermi surfaces are hole and electron pockets, respectively. **(c)** The evolution of the density of states (DOS) at the Fermi level as a function of pressure. Please note that a different setting of the elementary cell was used in a previous report[2].

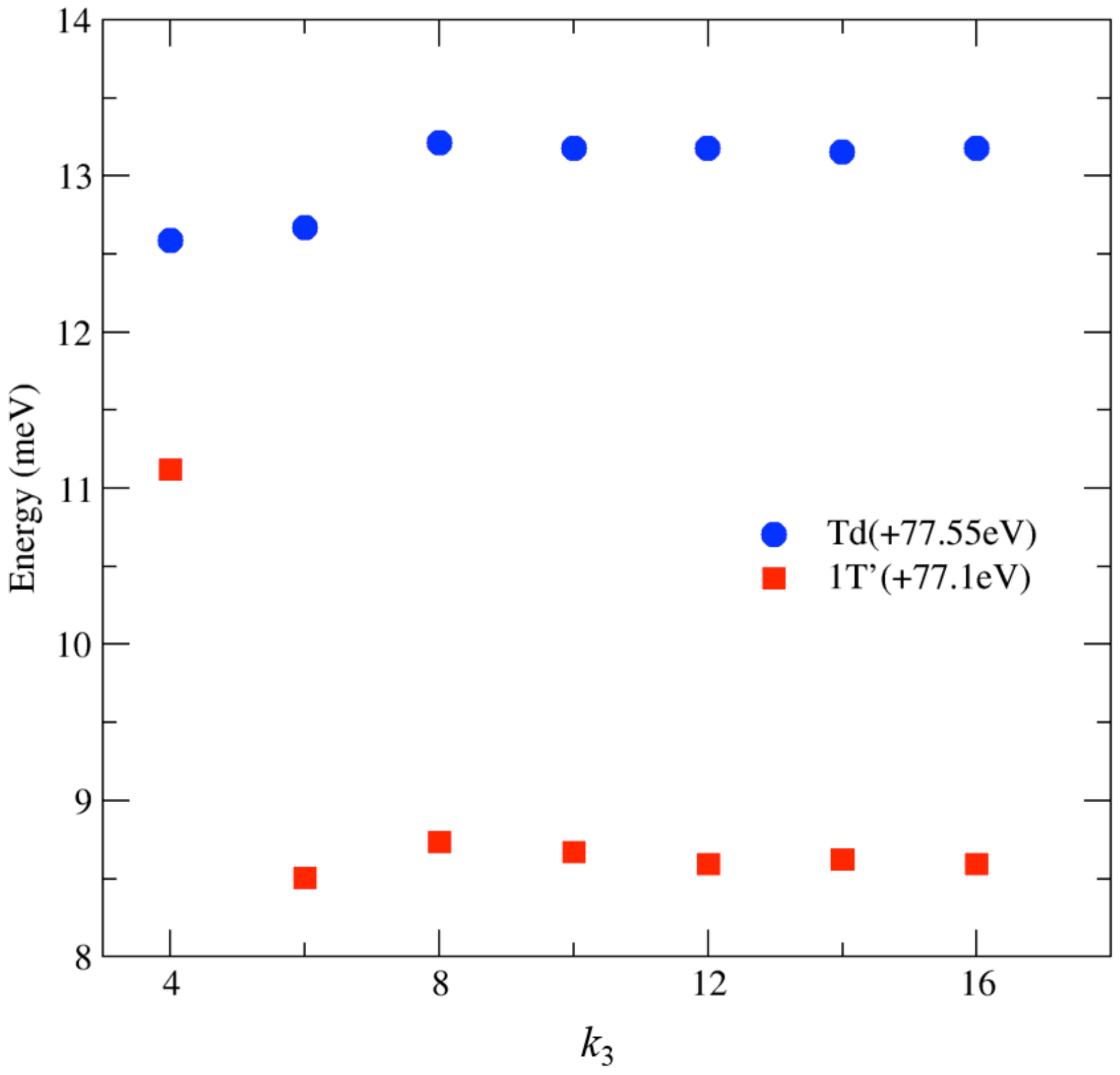

**Supplementary Figure 10. *k* grid convergence in DFT (GGA) calculations.** We use $k_3$ to represent the *k*-point density in reciprocal space. The *k* points in three directions have the relation of $3k_1=3k_2/2=k_3$. The convergence arrives at $k_3=8$.

## Supplementary Tables

| | |
|---|---|
| Phase | $T_d$-$MoTe_2$ ($\gamma$-$MoTe_2$) |
| Symmetry | Orthorhombic, *Pmn2₁* (No. 31) |
| Cell Parameters (Å) | $a$ = 3.477(2), $b$ = 6.335(3), $c$ = 13.889(6)<br>$\alpha = \beta = \gamma = 90^o$ |
| Wavelength (Å) | Mo $K\alpha$ - 0.71073 |
| $V$ (Å$^3$) | 305.97(20) |
| $Z$ | 4 |
| Calculated Density (g cm$^{-3}$) | 7.622(6) |
| Formula Weight (g mol$^{-1}$) | 351.14 |
| Absorption Coefficient (mm$^{-1}$) | 22.64 |
| $F_{000}$ | 584.0 |
| Reflections | 3328 |
| Data/Restraints/Parameters | 832 / 38 / 1 |
| Difference e-density (e/ Å$^3$) | +2.74 to -1.76 |
| $R_1$ (all reflections) | 0.0359 |
| $R_1$ $F$o > 2σ($F$o)) | 0.0312 |
| $wR_2$(all) | 0.076 |
| $R_{int}$/$R$(σ) | 0.0812/0.0192 |
| GooF | 1.044 |

**Supplementary Table 1. Single crystal structure determination for $T_d$ -$MoTe_2$ (γ-$MoTe_2$), experimental data taken at 120 K.**

| $T_d$-$MoTe_2$ | | | | | |
|---|---|---|---|---|---|
| Atom | Wyckoff Site | *x* | *y* | *z* | Occupation |
| Mo1 | 2*a* | 0 | 0.6052(2) | 0.5003(1) | 1 |
| Mo2 | 2*a* | 0 | 0.0301(2) | 0.0147(1) | 1 |
| Te1 | 2*a* | 0 | 0.8626(2) | 0.6557(1) | 1 |
| Te2 | 2*a* | 0 | 0.6405(2) | 0.1125(1) | 1 |
| Te3 | 2*a* | 0 | 0.2900(2) | 0.8593(1) | 1 |
| Te4 | 2*a* | 0 | 0.2160(2) | 0.4027(1) | 1 |

**Supplementary Table 2. Refined structural parameters for $T_d$-$MoTe_2$ at 120 K.**

| | Unit cell Volume ($Å^3$) | | Lattice parameters (Å) | | | | | | | |
|---|---|---|---|---|---|---|---|---|---|---|
| | Exp. | The. | ***a*** | | ***b*** | | ***c*** | | ***β*** | |
| | | | Exp. | The. | Exp. | The. | Exp. | The. | Exp. | The. |
| $T_d$ | 305.97 | 306.30 | 3.477 | 3.506 | 6.335 | 6.360 | 13.889 | 13.746 | 90 | 90 |
| 1T' | 303.64 | 306.48 | 3.469 | 3.495 | 6.330 | 6.392 | 13.860 | 13.741 | 93.917 | 93.8576 |

**Supplementary Table 3. Lattice parameters from experimental measurements (Exp.) and DFT relaxation (The.).**

**Supplementary Note 1**

We measured temperature dependent four-probe Hall resistivity in a field-sweep, $\rho_{xy}(H)$ to derive the Hall coefficient, $R_H(T)$, and type of carrier, where Hall coefficient directly gives the carrier concentration of a material. Importantly, $\rho_{xy}(H)$ exhibits a linear characteristic up to fields of 9 T. 1T'-$MoTe_2$ exhibits negative Hall coefficient, reflecting electrons as a major charge carriers in the whole temperature range. To calculate the electron charge density, $n_e$, we used the Drude single-band model[3], $n_e(T) = 1/[e\, R_H(T)]$, where $e$ is the elementary charge. The estimated values of $n_e$ are $5\times10^{19}$ $cm^{-3}$ at 2 K and $8\times10^{20}$ $cm^{-3}$ at 300 K (Supplementary Fig. 2), close to those reported [4].

Magnetoresistance (MR) is defined as the ratio of the change in resistivity due to the applied magnetic field, MR = $\rho_{xy}(H)/\rho_{xy}(0)$ -1. Here, we measured transverse MR (Supplementary Fig. 3). Resistivity of $T_d$-$MoTe_2$ is very sensitive to applied field and shows MR of $10^5$ % at 1.4 K without any sign of saturation up to $\mu_0 H$ = 66 T. More interestingly, these MR data display Shubnikov-de Haas (SdH) oscillations.

To calculate the amplitude of the SdH oscillations, we fit a 3rd order polynomial to the field dependent resistivity at each temperature. Supplementary Fig. 4a shows the fast Fourier transformation (FFT) of the SdH oscillations that exhibits a cyclotron frequency of the electrons at 245 T. This frequency is equivalent to the periodicity $1/B \approx 0.004$ $T^{-1}$ that corresponds to a cross-sectional area of the Fermi surface $A_F$ = 0.024 $Å^{-2}$ from the Onsager relationship $F = \Phi_0/(2\pi^2) A_F$, where $\Phi_0$ is the magnetic flux quantum. A very small Fermi momentum $k_F$ = 0.087 $Å^{-1}$ is obtained supposing a circular cross-section. Further, the cyclotron effective mass of the carriers is determined by fitting the following Lifshitz-Kosevich temperature reduction term[5] to the temperature dependent SdH oscillations amplitude (Supplementary Fig. 4b).

$$\frac{\Delta\rho_{xx}}{\rho_{xx}}(T) = \frac{14.69m^*T/B}{\sinh(14.69m^*T/B)}$$

The obtained effective mass, $m^*$ is 0.88 $m_0$, where $m_0$ is the bare mass of the electron. A lower $m^*$ is observed in few-layer 1T'-$MoTe_2$ specimens[3].

**Supplementary Note 2**

Consistent with previous reports[6], our calculations show that both $T_d$-$MoTe_2$ and 1T'-$MoTe_2$ are semimetals at zero pressure, as presented in Supplementary Fig. 9. Since the lattices for two phases are very close, their electronic structures are very similar to each other. As layered compounds, band dispersion is strong in the X-Γ-Y plane and weak along the Γ-Z direction. In the inset of Supplementary Fig. 9a and b, the Fermi surfaces are shown with hole and electron pockets located around the Γ point and the middle of Γ-Y, respectively. One main difference between them is the spin degeneracy. Due to the lack of inversion symmetry, band splitting occurs for spin-up and spin-down states in $T_d$-$MoTe_2$. In contrast, the 1T' phase has inversion symmetry and all the bands are spin degenerated. At the charge neutral point, we calculated the extreme Fermi surface area of the electron and hole pockets for both two phases. The electron pockets for $T_d$ phase are 0.021 and 0.019 $Å^{-2}$, which correspond to quantum oscillation frequencies of 221 and 200 T, respectively, according to the Onsager relation. Closely, the electron pocket in 1T' phase is 0.0169 $Å^{-2}$, corresponding to a quantum oscillation frequency of 178 T. The extreme Fermi surface area of hole pockets are much larger than 700 T. So we attribute the experimentally observed SdH oscillations frequency (Supplementary Fig. 4) to the electron pockets. The smaller theoretical frequency value, compared with experiment, may indicate that the sample is slightly electron-doped. Since the electron pocket areas are very close in the two phases, the measured quantum oscillations cannot distinguish them.

In Supplementary Fig. 6 the evolution of the lattice and electronic structures under pressure is shown. Due to van der Waals interactions, the lattice parameter *c* is compressed strongly by the applied pressure, while *a* and *b* only reduce slightly. When the compressive pressure increases from zero to 20 GPa, the density of states (DOS) at the Fermi energy first increases quickly and then decrease slightly after a maximum at around 10 GPa for both $T_d$ and 1T' phases (Supplementary Fig. 9c). This pressure dependence of DOS may be relevant to the dome-shape superconductivity observed in our experiment, since conventional superconductivity is known to be sensitive to the DOS at the Fermi energy.

**Supplementary References**

1. Alireza, P. L. & Lonzarich, G. G., Miniature anvil cell for high-pressure measurements in a commercial superconducting quantum interference device magnetometer. *Rev. Sci. Instrum.* **80**, 023906 (2009).

2. Sun, Y., Wu, S.-C., Ali, M. N., Felser, C., Yan, B. Prediction of the Weyl semimetal in the orthorhombic $MoTe_2$. *Phys. Rev. B* **92,** 161107 (2015).

3. Ashcroft, N. W. and Mermin, N. D. Solid State Physic (Harcourt, 1976).

4. Keum, D. H. *et al.* Bandgap opening in few-layered monoclinic $MoTe_2$. *Nature Phys.* **11**, 482–486 (2015).

5. Shoenberg, D. Magnetic Oscillations in Metals. Cambridge University Press (2009).

6. Dawson, W. G. & Bullett, D. W. Electronic structure and crystallogrophy of $MoTe_2$ and $WTe_2$. *J. Phys. C: Solid State Phys.* **20,** 6159-6174 (1987).